\documentclass[twocolumn,prc,superscriptaddress,amsmath,amssymb]{revtex4-2}

\usepackage{amsmath,amsthm,amssymb}
\usepackage{graphicx}
\usepackage{booktabs}
\usepackage{color}
\usepackage{xcolor}
\usepackage{float}
\usepackage{multirow,array}

\usepackage[colorlinks,
bookmarks=true,
linkcolor=blue,
urlcolor=blue,
anchorcolor=black,
citecolor=blue
]{hyperref}

\colorlet{darkblue}{blue!60!black}

\begin{document}

\title{Elliptic flow of $\phi$ and $\Omega$ via quark recombination mechanism in relativistic heavy-ion collisions}


\author{Zhidong Yang}
\email{zdyang07@163.com}
\affiliation{School of Physics, Hubei University, Wuhan, Hubei 430062, China}

\author{Lie-Wen Chen}
\email{lwchen@sjtu.edu.cn}
\affiliation{State Key Laboratory of Dark Matter Physics, Key Laboratory for Particle Astrophysics and Cosmology (MOE), and Shanghai Key Laboratory for Particle Physics and Cosmology, School of Physics and Astronomy, Shanghai Jiao Tong University, Shanghai 200240, China}

\date{\today}

\begin{abstract}

Multi-strange hadrons, such as the $\phi$ meson and $\Omega$ baryon, serve as unique probes for the properties of the quark-gluon plasma in relativistic heavy-ion collisions. Using a quark recombination model with the quark phase-space information parameterized in a viscous blastwave, we show that the elliptic flow $v_2$ of $\phi$ and $\Omega$ can be describled by quark recombination in a wide energy range from $\sqrt{s_{NN}}$ = 7.7 GeV to 5.02 TeV. We check the scaling properties of $v_2$ of $\phi$ and $\Omega$ and find the number-of-constituent-quark (NCQ) scaled $v_2$ of $\phi$ and $\Omega$ are approximately equal. We further present the $v_2$ of strange quarks at hadronization extracted from $\phi$ and $\Omega$ data. We find the $v_2$ of strange quarks is close to the NCQ-scaled $v_2$ of $\phi$ and $\Omega$ at $\sqrt{s_{NN}}$ = 7.7-200 GeV, but significant larger than the latter at $\sqrt{s_{NN}}$ = 2.76-5.02 TeV, suggesting that the $v_2$ of strange quarks cannot be simply equated to the NCQ-scaled $v_2$ of $\phi$ and $\Omega$. Our results demonstrate that the quark recombination mechanism is valid for $\phi$ and $\Omega$ production in heavy-ion collisions at $\sqrt{s_{NN}}$ = 7.7 GeV to 5.02 TeV,  indicating the existence of constituent quark degree of freedom at these energies.

\end{abstract}



\maketitle

\section{Introduction}

Lattice Quantum Chromodynamics (LQCD) calculations predict a transition from ordinary nuclear matter to a new state of matter that consist of deconfined quarks and gluons, called quark gluon plasma (QGP) \cite{Fodor:2004nz}. The existence of the QGP has been supported by many observations of ultra-relativistic heavy-ion collisions at the Relativistic Heavy Ion Collider (RHIC) and the Large Hadron Collider (LHC) \cite{STAR:2005gfr,PHENIX:2004vcz,Muller:2012zq}. One important observable is the hadronic collective anisotropic flow, such as elliptic flow $v_2$, the second order Fourier coefficient of the azimuthal particle spectrum in momentum space, which is closely related to early stage dynamics of heavy-ion collisions \cite{Ollitrault:1992bk,Sorge:1998mk}. Early experimental results from RHIC \cite{STAR:2003wqp, STAR:2005npq, PHENIX:2006dpn} found that at intermediate transverse momenta ($p_T$= 2-5 GeV/$c$), the $v_2$ of mesons and baryons approximately fall on a universal curve when both $v_2$ and $p_T$ are divided by number of constituent quarks ($n_q$). The $n_q$-scaled $v_2$ of hadrons is usually regarded as the elliptic flow of quarks \cite{Molnar:2003ff, Kolb:2004gi}. The number-of-constituent-quark (NCQ) scaling can then be expressed as 
\begin{equation}
	v_{2,h}(p_T) \approx n_q v_{2,q}(p_T/n_q)
	\label{eq:ncq}
\end{equation}
where $n_q=2$ (3) for mesons (baryons), $v_{2,h}$ and $v_{2,q}$ are the elliptic flows of hadrons and quarks respectively. The NCQ scaling provides strong evidence for quarklike degrees of freedom and collective flow at the parton level, and is usually regarded as a clear signature for the creation of QGP.

The NCQ scaling can be well explained by quark recombination or coalescence models \cite{Greco:2003xt,Greco:2003mm,Fries:2003kq,Fries:2003vb}, indicating the quark recombination mechanism plays an important role in hadron production in the intermediate-$p_T$ region. However, later results from the RHIC found the scaling property dependents on the collision centrality \cite{PHENIX:2012swz}. They observed good consistency with NCQ scaling for the elliptic flow $v_2$ of pions, kaons and protons in central collisions, but significant deviations from this scaling behavior in the non-central collisions at $\sqrt{s_{NN}}$ = 200 GeV. Experiments at the LHC also reported about 20\% deviation from NCQ scaling for the elliptic flow $v_2$ of identified hadrons when $p_T>3$ GeV/$c$ at $\sqrt{s_{NN}}$ = 2.76 TeV \cite{ALICE:2014wao}. This observation has sparked theoretical concerns regarding the NCQ scaling behavior \cite{Singha:2016aim,Zheng:2016iia,Liu:2025eml}. Since NCQ scaling is conventionally taken as a necessary feature of quark recombination, its experimental violation may challenge this mechanism.

‌Therefore, it is worthwhile to investigate the NCQ scaling behavior across different collision centralities and collision energies. For this purpose, we perform a systematic analysis on the experimental data of $\phi$ meson and $\Omega$ baryon at $\sqrt{s_{NN}}$ = 7.7-200 GeV at the RHIC and $\sqrt{s_{NN}}$=2.76-5.02 TeV at the LHC, utilizing a quark recombination model developed in Ref.\ \cite{Yang:2022ixy}. In the recombination model, the quark phase-space information is parameterized in a viscous blastwave, which includes non-equilibrium deformations of thermal distributions due to shear and bulk stresses. By introducing viscous corrections, we naturally account for the observed saturation feature of elliptic flow $v_2$ at intermediate transverse momenta. 

The $\phi$ and $\Omega$ have relatively small hadronic interaction cross sections~\cite{Shor:1984ui}, and thus carry direct information from early stage with negligible hadronic effects~\cite{Shor:1984ui,VanHecke:1998yu,Chen:2005nw,Chen:2006vc,Chen:2008vr,Auvinen:2016uuv,Hwa:2016qtb,Ye:2017ooc,Pu:2018eei,Song:2019sez}. In addition, $\phi(s\bar{s})$ and $\Omega^-(sss)$ ( $\bar{\Omega}^+(\bar{s}\bar{s}\bar{s})$) are both purely made up of (anti-)strange quarks, so that the NCQ scaling, assuming its existence, would appear in its simplest form, making it easy to extract the $v_2$ of strange quarks at hadronization. Furthermore, Eq.(\ref{eq:ncq}) is a direct result of the naive quark coalescence model, which holds only at low quark phase-space density and is gradually broken at high density \cite{Molnar:2003ff}. Since strange quarks have much smaller phase-space density than up/down quarks (with protons, kaons and pions dominating most bulk observables), they are expected to better satisfy the NCQ scaling.

In this work, we show that the elliptic flow $v_2$ of $\phi$ and $\Omega$ can be describled by quark recombination in a wide energy range from $\sqrt{s_{NN}}$ = 7.7 GeV to 5.02 TeV. We check the scaling properties of $v_2$ of $\phi$ and $\Omega$ and find the NCQ-scaled $v_2$ of $\phi$ and $\Omega$ are approximately equal. We further present the $v_2$ of strange quarks at hadronization extracted from $\phi$ and $\Omega$ data. We find the $v_2$ of strange quarks is close to the NCQ-scaled $v_2$ of $\phi$ and $\Omega$ at $\sqrt{s_{NN}}$ = 7.7-200 GeV, but significant larger than the latter at $\sqrt{s_{NN}}$ = 2.76-5.02 TeV, suggesting that the $v_2$ of strange quarks cannot be simply equated to the NCQ-scaled $v_2$ of $\phi$ and $\Omega$, and special cautions are required when extracting the elliptic flow of partons from the measured elliptic flow of hadrons. The paper is organized as follows: in section \ref{sec:qr}, we introduce the quark recombination model, as well as viscous blastwave parameterization for quark phase-space distribution. In section \ref{sec:results} we present and discuss our results. In section \ref{sec:sum} we summarize our conclusions.

\section{Model and Methods}
\label{sec:qr}

\subsection{Theoretical model}\label{subsec:model}

Quark recombination or coalescence models were initially proposed to explain the baryon-over-meson enhancement and number-of-constituent-quark scaling observed in RHIC Au+Au collisions~\cite{Greco:2003xt,Greco:2003mm,Fries:2003vb,Fries:2003kq,Hwa:2002tu}.
In a recent work \cite{Yang:2022ixy}, we introduce viscous corrections into quark recombination, and here give a brief overview of the formalism. In the following, $\mathbf{r}$ and $\mathbf{p}$ are the 3-position and 3-momentum respectively, $m$ is particle mass and $p^\mu = (E,~\mathbf{p})$ is the 4-momentum of hadrons with $E=\sqrt{m^2+p^2}$. Following Refs.~\cite{Fries:2003kq,Fries:2003vb,Fries:2008hs}, the momentum distribution of mesons is given by
\begin{eqnarray}
	E\frac{dN_M}{d^3\bf p}&=& C_M\int_\sigma \frac{p^{\mu}\cdot d\mathbf{\sigma_\mu}}{(2\pi)^3}
	\int dx_1 dx_2 \Phi_M(x_1,x_2) \nonumber\\ &&\times \gamma_a \gamma_b f_a({\bf r},x_1\mathbf{p})f_b({\bf r},x_2\mathbf{p}).\
	\label{eq:meson}
\end{eqnarray} 
where $C_M$ is the spin degeneracy factor of a given meson species, $\sigma$ is the hypersurface of hadronization, $\varPhi_M$ is the effective wave function squared of mesons, $x_{1,2}$ are light cone coordinates defined as ${\bf p}_{1,2}=x_{1,2}\bf p$, $\gamma_{a,b}$ are fugacity factors for quarks, and $f_{a,b}$ are the quark phase-space distributions. The $\varPhi_M$ is parameterized as a Gaussian distribution
\begin{equation}
   \Phi_M=\frac{2}{\sqrt{2\pi}\sigma_M} \exp (-\frac{(x_1-x_a)^2+(x_2-x_b)^2}{\sigma_M^2})\delta(x_1+x_2-1)
   \label{eq:wave}
\end{equation}
where $\sigma_M$ is the variance, $x_{a,b}=m_{1,2}/(m_1+m_2)$ are the peak values, and $m_{1,2}$ are the masses of the constituent quarks. With this treatment, we ensure that $\varPhi_M$ reaches its maximum value when the average velocity of the constituent quarks is identical. Similar expression can be derived for baryons.

The quark phase-space distribution is parameterized in a viscous blastwave~\cite{Yang:2016rnw,Yang:2020oig,Yang:2022yxa}, based on the Retiere and Lisa blastwave \cite{Retiere:2003kf}. The quark distribution is given by
\begin{equation}
	f(r,p) = f_{0}(r,p) + \delta f_{\rm {shear}}(r,p)+\delta f_{\rm{bulk}}(r,p)
	\label{eq:fdist}
\end{equation}
where $f_0(r,p) = 1/(e^{(u^\mu p_{\mu}-\mu_i)/T}\mp 1)$ is the equilibrium Bose/Fermi distribution as a function of the flow field $u^\mu$, particle momentum $p^\mu$, the local temperature $T$ and chemical potentials $\mu_i=b_i\mu_B + s_i\mu_S + q_i\mu_Q$ with baryon number $b_i$, strangeness $s_i$ and electric charge $q_i$ (we assume $\mu_Q=0$ in this work). $\delta f_{\rm {shear}}$ and $\delta f_{\rm{bulk}}$ denote corrections from the shear and bulk viscosities, respectively. For the shear viscous corrections, we use the Grad's method \cite{Song:2007ux,Damodaran:2020qxx}
\begin{equation}
	\delta f_{\rm {shear}} = \frac{1}{2T^2} \frac{p_\mu p_\nu}{\epsilon+P}\pi^{\mu\nu} f_{0}(1\pm f_{0})
	\label{eq:shear}
\end{equation}
where $\epsilon$ is the energy density, $P$ is the pressure, $\pi^{\mu\nu}$ is the shear stress tensor and $+(-)$ for bosons (fermions). In the Navier-Stokes approximation $\pi^{\mu\nu}=2\eta \sigma^{\mu\nu}$ where $\eta$ is the shear viscosity and $\sigma^{\mu\nu}$ is the shear gradient tensor defined as $
\sigma^{\mu\nu} = \frac{1}{2} \left( \nabla^\mu u^\nu + \nabla^\nu u^\mu \right) - \frac{1}{3} \Delta^{\mu\nu} \nabla_\lambda u^\lambda$ with flow field $u^\mu$, $\nabla^\mu = \Delta^{\mu\nu} \partial_\nu$ and $\Delta^{\mu\nu} = g^{\mu\nu} - u^\mu u^\nu$. The spatial derivatives in  $\sigma^{\mu\nu}$ can be obtained directly. The time derivatives cannot be given by blastwave itself, and are determined by solving ideal hydrodynamics~\cite{Yang:2022yxa}.

For the bulk viscous corrections, we use the 14-moment approximation \cite{Denicol:2014vaa,Paquet:2015lta}
\begin{equation}
	\delta f_{\rm bulk}=- f_0(1\pm f_0) \Pi \frac{\tau_\Pi}{\zeta} \left[ \frac{1}{3} \frac{m^2}{T} \frac{1}{p^\mu u_\mu}+\frac{p^\mu u_\mu}{T} \left(c_s^2-\frac{1}{3} \right) \right]
	\label{eq:bulk1}
\end{equation}
where $\zeta$ is the bulk viscosity, $\Pi$ is the bulk viscous pressure, $\tau_\Pi$ is the bulk relaxation time and  $c_{s}^2$ is the velocity of sound squared. At the first order approximation, one has $\Pi=-\zeta \partial_\mu u^\mu$. Following Ref.\ \cite{Denicol:2014vaa}, the ratio of bulk viscosity to bulk relaxation time is given by
\begin{equation}
	\frac{\zeta }{\tau _{\Pi }}=\left( \frac{1}{3}
	-c_{s}^{2}\right) \left( \varepsilon +P \right) -\frac{2}{9}\left(
	\varepsilon-3P\right) -\frac{m^{4}}{9}I_{-2,0}
\end{equation}
with $I_{-2,0}$ defined as $I_{-2,0}=\frac{g}{(2\pi)^3}\int \frac{d^3 \mathbf{p}}{E} \frac{f_{0}(r,p)}{ (p^{\mu} u_{\mu})^2}$ and $g$ is the degeneracy factor.

Let us now describe the blastwave parameterization for the flow field $u^\mu$. Here $R_{x,y}$ are the semi axes of the fireball at freeze-out, $\rho=\sqrt{x^2/R_x^2+y^2/R_y^2}$ is the reduced radius, $\phi_s=\arctan y/x$ is the azimuthal angle, and $\eta_s = \frac{1}{2}\ln\frac{t+z}{t-z}$ is the space-time rapidity. The hadronization hypersurface is assumed to be constant $\tau=\sqrt{t^2-z^2}$, which approximates hadron formation as an instantaneous process on a thin space-time hypersurface. The flow field is parameterized as
\begin{eqnarray}
	u^{\mu}&=&(\cosh\eta_s\cosh\eta_T,\sinh\eta_T\cos\phi_b, \nonumber\\ && \qquad \sinh\eta_T\sin\phi_b,\sinh\eta_s\cosh\eta_T)
	\label{eq:flow}
\end{eqnarray} where $\eta_T$ is the transverse flow rapidity and $\phi_b$ is the azimuthal angle of $u^\mu$ in the transverse plane. $\eta_T$ is given by the transverse velocity $v_T=\tanh\eta_T$ with
\begin{equation}
	v_T=\rho^n (\alpha_0+\alpha_2\cos2\phi_b)
\end{equation}
where $\alpha_0$ is the average surface velocity, $\alpha_2$ is an elliptic deformation of the flow field and $n$ is a power term. In this work, we use a linear expression and set $n=1$. The transverse flow vector is chosen to be perpendicular to the elliptic surface at $\rho=1$. The ratio $R_y/R_x$ significantly influences elliptic flow, so we choose $R_y/R_x$ as a fit parameter and constrain $R_x$, $R_y$ and $\tau$ by adding a simple geometric estimate $R_x \approx (R_0-b/2)+ 0.65 \tau (\alpha_0+\alpha_2)$ where $R_0$ is the radius of the colliding nucleus and $b$ is the impact parameter. The values of $b$ used for each centrality bin are based on Glauber Monte Carlo simulations for related experiments \cite{ALICE:2013hur,HADES:2017def}. The information of the quark phase-space freeze-out configuration at the moment of hadronization are thus determined by 7 parameters: $\tau, T, \alpha_0, \alpha_2, R_y/R_x, \eta T/(\epsilon + P), \zeta T/(\epsilon + P)$.

\subsection{Experimental data and fit parameters}

In order to constrain the unknown parameters embedded in the quark recombination model, we utilize the transverse-momentum~($p_T$) spectra and elliptic flows $v_2$ of $\phi$ and $\Omega$ as our observables. We will only consider mid-rapidity ($y=0$) here. The data we used are from the STAR collaboration for Au+Au collisions at $\sqrt{s_{NN}}$ = 7.7, 11.5, 19.6, 27, 39, 54.4, 62.4 and 200 GeV~\cite{STAR:2006egk,STAR:2007mum,STAR:2008bgi,STAR:2010yyv,STAR:2015gge,STAR:2015rxv,STAR:2015vvs,STAR:2019bjj,STAR:2022tfp,Liu:2022mxa,Dixit:2022geb} (STAR Preliminary data for $v_2$ at 19.6 GeV), and the ALICE collaboration for Pb+Pb collisions at $\sqrt{s_{NN}}$ = 2.76, 5.02 TeV~\cite{ALICE:2013xmt,ALICE:2014wao,ALICE:2017ban, ALICE:2018yph,ALICE:2019xyr,ALICE:2022zks} around mid-rapidity, as listed in Tab.\ \ref{tab:data}. For Au+Au collisions at $\sqrt{s_{NN}}$ = 7.7-62.4 GeV, the elliptic flow $v_2$ data are selected in the 10–40\% centrality bin. This specific centrality range is chosen to minimize the contributions from non-flow correlations \cite{Goudarzi:2020eoh}. Non-flow correlations are reaction-plane-independent azimuthal correlations, arising from transverse momentum conservation, resonance decays, inter/intra- jet correlations, etc.\  \cite{Borghini:2000cm,STAR:2004jwm}. Non-flow contributions can bias the experimentally measured anisotropic flow, and most of them scale with event multiplicity N as 1/N \cite{Borghini:2000cm}. In 0–10\% central collisions, small $v_2$ is dominated by initial state fluctuations; in 40–80\% peripheral collisions, low multiplicity enhances non-flow effects. As a result, non-flow interference is minimized in the 10–40\% centrality.

\begin{table}[tbh]
	\caption{\label{tab:data} Experimental data of $\phi$ and $\Omega$ used in our analysis.}
	\centering
		\begin{tabular}{l|l|l}
			\hline
			$\sqrt{s_{NN}}$(GeV) & \multicolumn{2}{c}{Centrality}  \\
			\hline
			\multirow{2}*{7.7, 11.5} & $v_2(p_T)$ & 10-40\% \cite{STAR:2015rxv}   \\
			    & $\frac{dN}{d^2p_T dy} (\phi)$   & 20-30\% \cite{STAR:2015vvs,STAR:2019bjj} \\
			    & $\frac{dN}{d^2p_T dy} (\Omega)$ & 0-60\% (10-60\%) \cite{STAR:2015vvs,STAR:2019bjj} \\
			\hline
			\multirow{2}*{19.6, 27, 39} & $v_2(p_T)$ & 10-40\% \cite{STAR:2015rxv,Liu:2022mxa,Dixit:2022geb} \\ 
                & $\frac{dN}{d^2p_T dy} (\phi)$   & 20-30\%  \cite{STAR:2015vvs,STAR:2019bjj} \\
                & $\frac{dN}{d^2p_T dy} (\Omega)$ & 20-40\% \cite{STAR:2015vvs,STAR:2019bjj} \\
			\hline
			\multirow{2}*{54.4, 62.4} & $v_2(p_T)$ & 10-40\% \cite{STAR:2015rxv,STAR:2022tfp} \\
		     	& $\frac{dN}{d^2p_T dy}$ & 20-40\% \cite{STAR:2008bgi,STAR:2010yyv} \\
			\hline
			\multirow{3}*{200} & $v_2(p_T)$  &  \enspace0-30\%, 30-80\% \cite{STAR:2015gge} \\
			& $\frac{dN}{d^2p_T dy}(\phi)$   &  10-20\%, 40-50\%  \cite{STAR:2007mum}\\
			& $\frac{dN}{d^2p_T dy}(\Omega)$ &  \quad 0-5\%, 40-60\% \cite{STAR:2006egk}\\
			\hline
			\multirow{3}*{2760, 5020} & $v_2(p_T)$ 	&\multirow{2}*{10-20\%, 20-30\%, 30-40\%,	40-50\%  \cite{ALICE:2014wao,ALICE:2017ban}}  \\
			& $\frac{dN}{d^2p_T dy}(\phi)$                      \\
			& $\frac{dN}{d^2p_T dy} (\Omega)$ & 10-20\%, 20-40\%, 40-60\% \cite{ALICE:2013xmt} \\
			\hline
	\end{tabular}
\end{table}

Due to data availability, the centrality bins for spectra are slightly different from that of $v_2$, as shown in Tab.\ \ref{tab:data}. For Au+Au collisions at 7.7–62.4 GeV, we adopt the experimental data of $\Omega^-$. For collisions at 200 GeV and 2.76 TeV, we use the combined dataset of $\Omega^-$ and $\bar{\Omega}^+$. For the $\phi$ and $\Omega$ spectra at 54.4 GeV, the corresponding measurements are not available yet, so we take the spectra at 62.4 GeV as a reasonable substitute. As for the spectra at 5.02 TeV, no published experimental data for $\Omega$ is currently accessible, hence only the spectra of $\phi$ are included in our analysis.

For the $\phi$ meson, the fitted $p_T$ range is $p_T<3$ GeV/$c$ at collision energies of 7.7–62.4 GeV, and $p_T<3.5$ GeV/c at 200 GeV, 2.76–5.02 TeV, with slight adjustments due to limited experimental data. For the $\Omega$ baryon, the fitted $p_T$ range is $p_T<3.8$ GeV/$c$ at 7.7–62.4 GeV, and $p_T<4.5$ GeV/c at 200 GeV, 2.76–5.02 TeV. We have checked that the correction $\delta f(r,p)$ is small for both $\phi$ and $\Omega$ within the above fit ranges, i.e., less than $30\%$ of $f_0$ for the majority of points, which is much smaller than the commonly adopted upper bound $\approx 1$~\cite{JETSCAPE:2020mzn,Teaney:2003kp}, ensuring the applicability of the viscous corrections.

\begin{table}[tb]
	\caption{\label{tab:paraset} Values of temperature $T$, $\eta T/(\epsilon + P)$ and baryon chemical potential $\mu_B$ used in quark recombination for different collision energies. A constant $\zeta T/(\epsilon + P)=0.03$ is used for all collision energies.}
	\centering
	\resizebox{\columnwidth}{!}{
	\begin{tabular}{l|l|llllllll} 
		\hline
		\multicolumn{2}{c|}{$\sqrt{s_{NN}}$ (GeV)} &7.7 &11.5 &19.6 &27  & 39 &54.4 &62.4 & $\ge 200$ \\
		\hline
		\multicolumn{2}{c|}{$T$ (MeV)}  & 145 & 150 & 155  & 157.5 & 157.5 & 158.5 & 158.5 & 160  \\
		\hline
		\multicolumn{2}{c|}{$\eta T/(\epsilon + P)$}& 0.08 & 0.08  & 0.08  &0.10 & 0.11 & 0.14 & 0.15 & 0.20 \\
		\hline
		\multicolumn{2}{c|}{$\mu_B$ (MeV) } & 406 & 303 & 197 & 149 & 107  & 79 & 69 & 0 \\
		\hline
	\end{tabular}	
    }
\end{table}

Now we discuss the fit parameters used in the model. Since the primary goal of this work is to investigate the scaling behaviors of elliptic flow $v_2$, we do not perform a precise tuning on several non-critical parameters. Instead, these parameters are fixed at physically reasonable values, as long as the final physical conclusions are not sensitive to their variations. For example, strange quark mass variation significantly changes the yields of $\phi$ and $\Omega$, but barely affects their elliptic flow $v_2$ \cite{Chen:2006vc}. A similar behavior also applies to the hadron wave function variances $\sigma_M$ and $\sigma_B$. In this work, we specify the strange quark mass as $m_s=500$ MeV and hadron wave function variances as $\sigma_M=0.3$, $\sigma_B=0.1$. We have confirmed that the variations in $m_s$, $\sigma_M$ and $\sigma_B$ have minimal impact on our results.

The temperature $T$ and baryon chemical potential $\mu_B$ at hadronization are taken from Ref.\ \cite{Andronic:2017pug}, by assuming their values are close to the corresponding values at chemical freeze-out for different collision energies. The strangeness chemical potential $\mu_S$ is considerably smaller than $\mu_B$ \cite{STAR:2019bjj} and is thus fixed at $\mu_S=0$ in our calculation. The quantities $\eta T/(\epsilon + P)$ and $\zeta T/(\epsilon + P)$, which reduce to specific shear and bulk viscosities $\eta/s$ and $\zeta/s$ when $\mu_B=0$ (we have $\epsilon + P=Ts+\mu_B n_B$), depend on temperature and baryon chemical potential $\mu_B$ and their exact values are still under active exploration \cite{Karpenko:2015xea,Auvinen:2017fjw,Gotz:2022naz,Shen:2023awv,Yang:2023apw}. Based on the results reported in Ref.\ \cite{Yang:2023apw}, the values of $\eta T/(\epsilon + P)$ are listed in Tab.\ \ref{tab:paraset}, while $\zeta T/(\epsilon + P)$ is fixed to a constant $\zeta T/(\epsilon + P)=0.03$ for all collision energies. The full set of $(T, \mu_B, \eta T/(\epsilon + P))$ used in our calculation are listed in Tab.\ \ref{tab:paraset}. 

\begin{table}[tb]
	\caption{\label{tab:fit} Freeze-out parameters of strange quarks at hadronization in Au+Au collisions at $\sqrt{s_{NN}}$ = 7.7-200 GeV and Pb+Pb collisions at $\sqrt{s_{NN}}$ =2.76-5.02 TeV. The corresponding values of $T$, $\eta T/(\epsilon + P)$ and $\zeta T/(\epsilon + P)$ are given in Tab.\ \ref{tab:paraset}.}
	\centering
	\begin{tabular}{lllllll} 
		\hline
		$\sqrt{s_{NN}}$ & Centrality &$\tau$ (fm/$c$) & $\alpha_0(c)$ &$\alpha_2(c)$  & $R_y/R_x$  \\
		\hline
		7.7 GeV  &10-40\% & 6.3  & 0.473 & 0.031  & 1.20  \\
		11.5 GeV &        & 6.7  & 0.509 & 0.030  & 1.23  \\
		19.6 GeV &        & 6.9  & 0.521 & 0.023  & 1.28  \\
		27 GeV   &        & 6.8  & 0.575 & 0.025  & 1.22  \\
		39 GeV   &        & 6.9  & 0.603 & 0.028  & 1.22  \\
		54.4 GeV &        & 6.9  & 0.635 & 0.031  & 1.24  \\
		62.4 GeV &        & 6.8  & 0.641 & 0.031  & 1.25  \\
		\hline
		200 GeV  &0-30\%  & 7.9  & 0.717 & 0.027  & 1.14  \\
		         &30-80\% & 5.1  & 0.692 & 0.051  & 1.35  \\
		\hline
		2.76 TeV &10-20\% & 9.6  & 0.820 & 0.022  & 1.21  \\
		         &20-30\% & 9.3  & 0.808 & 0.034  & 1.23  \\
	   	         &30-40\% & 7.8  & 0.811 & 0.030  & 1.34  \\
		         &40-50\% & 6.8  & 0.794 & 0.034  & 1.44   \\
		\hline
		5.02 TeV &10-20\% & 10.0  & 0.843 & 0.014  & 1.27  \\
		         &20-30\% & 9.4  & 0.835 & 0.024  & 1.30  \\
		         &30-40\% & 8.3  & 0.827 & 0.029  & 1.38  \\
		         &40-50\% & 7.2  & 0.821 & 0.030  & 1.45   \\
		\hline
	\end{tabular}	
\end{table}

For the fugacity factor (see Eq.\ (\ref{eq:meson})), we set $\gamma_s=0.65$ for 7.7-62.4 GeV and $\gamma_s=0.8$ for 200 GeV, 2.76-5.02 TeV. The same values are used for $\gamma_{\bar{s}}$. Note that the value of $\gamma_s$ for 7.7-62.4 GeV are smaller than those obtained in thermal models~\cite{STAR:2010yyv,STAR:2017sal}. The value of $\gamma_s$ for 200 GeV, 2.76-5.02 TeV is derived from quark coalescence model~\cite{Pu:2018eei}. We have checked the impact of $\gamma_s$ and observed minor effects on the final results. Regarding other constants, we use sound speed squared $c^2_s=0.15$ (see Eq.\ (\ref{eq:bulk1})) for the QGP at hadronization \cite{Huovinen:2009yb,Monnai:2019hkn}, spin degeneracy factor $C_M=3$ for $\mathrm{\phi}$ and $C_B=4$ for $\Omega^-$.

\begin{figure}[tb]
	\centering
	\begin{minipage}{0.48 \textwidth}
		\includegraphics[width=1\textwidth]{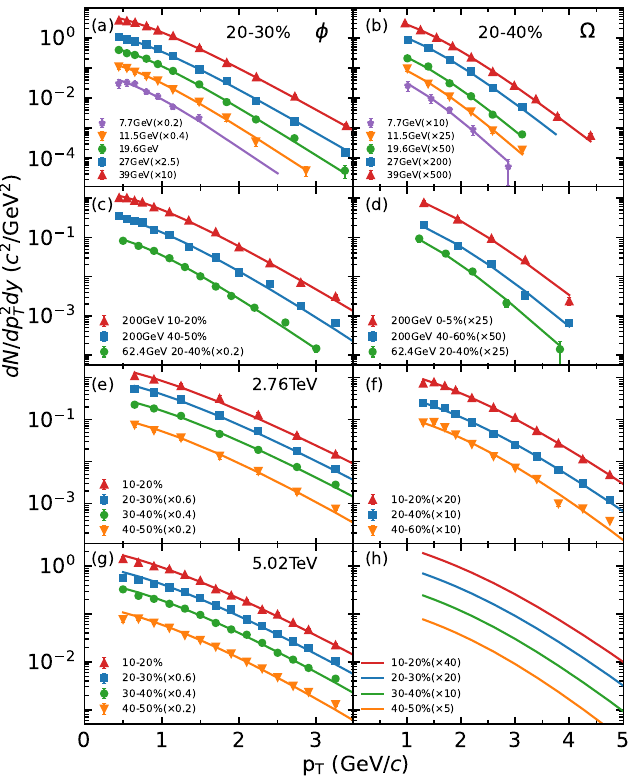}
	\end{minipage}
	\caption{\label{fig:allsp}
		The calculated (lines) and measured (symbols) transverse-momentum spectra of $\phi$ mesons (a,c,e,g) and $\Omega$ baryons (b,d,f,h) at mid-rapidity in Au+Au collisions at $\sqrt{s_{NN}}$ = 19.6-200 GeV and Pb+Pb collisions at $\sqrt{s_{NN}}$ = 2.76-5.02 TeV. Solid lines are quark recombination calculation. Symbols are data from STAR and ALICE measurement.}	
\end{figure}

The parameters left in our model are ($\tau$, $\alpha_0$, $\alpha_2$, $R_y/R_x$), namely hadronization proper time $\tau$, surface radical velocity $\alpha_0$, velocity deformation $\alpha_2$, ratio of event plane radii $R_y/R_x$. These remaining parameters determine the quark phase-space freeze-out properties at hadronization. For each centrality bin at each collision energy, the fluid has unique values for ($\tau$, $\alpha_0$, $\alpha_2$, $R_y/R_x$), which can be extracted from the experimental data.

\begin{figure*}[tb]
	\centering
	\includegraphics[width=0.93\textwidth]{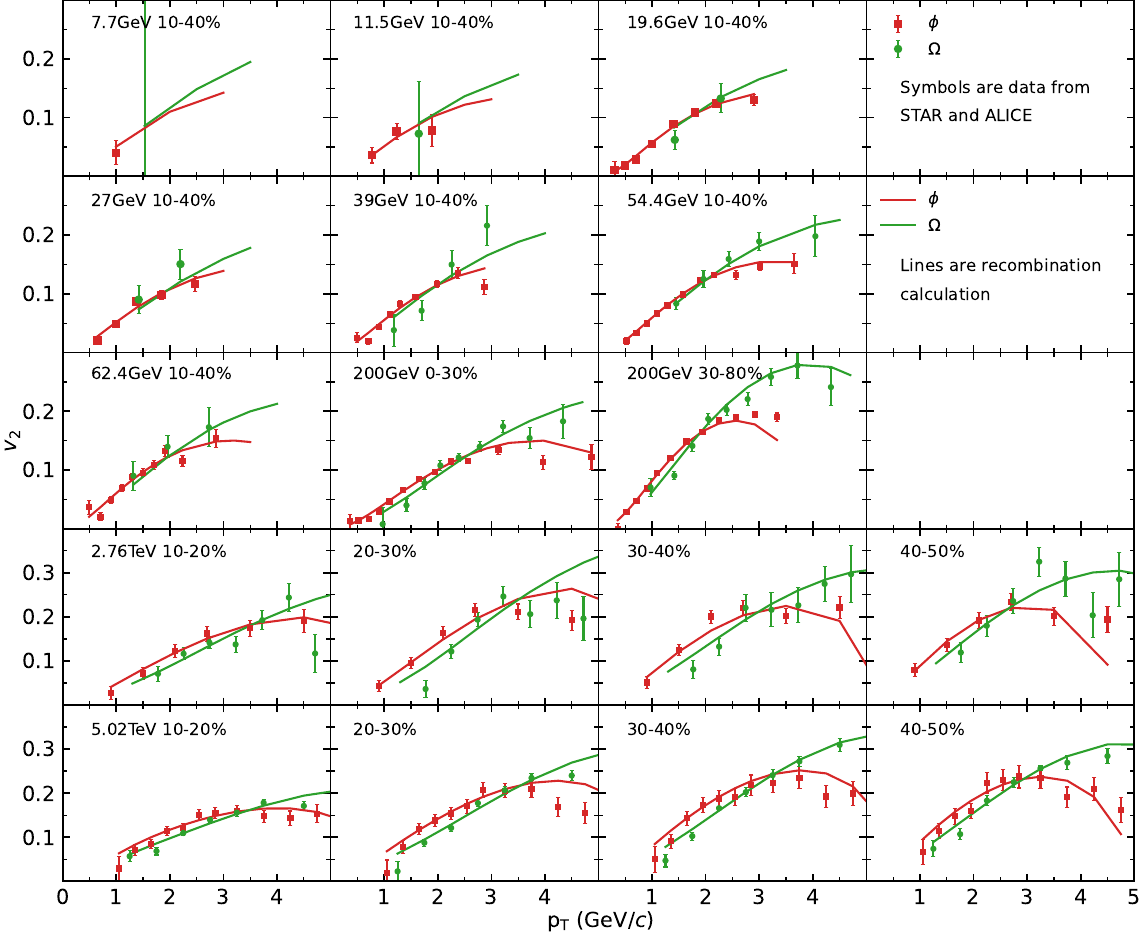}
	\caption{\label{fig:allv2}
		The calculated (lines) and measured (symbols) elliptic flows $v_2$ of $\phi$ and $\Omega$ at mid-rapidity in Au+Au collisions at $\sqrt{s_{NN}}$ = 7.7-200 GeV and Pb+Pb collisions at $\sqrt{s_{NN}}$ = 2.76-5.02 TeV. Solid lines are quark recombination calculation, and symbols are data from STAR and ALICE measurement.}	
\end{figure*}

\section{fit results and discussion}
\label{sec:results}

By fitting the $p_T$-differential spectra and elliptic flows $v_2$ of $\phi$ mesons and $\Omega$ baryons, we extract the strange quark freeze-out configurations, as summarized in Table.\ \ref{tab:fit}. Within our quark recombination framework, we compute the $p_T$ spectra and elliptic flows $v_2$ of $\phi$ and $\Omega$, and confront the model outputs with the experimental data. Fig.\ \ref{fig:allsp} and Fig.\ \ref{fig:allv2} show our results for $p_T$ spectra and elliptic flow $v_2$ of $\phi$ and $\Omega$ in RHIC Au+Au collisions and LHC Pb+Pb collisions. As one can see from Fig.\ \ref{fig:allsp} and Fig.\ \ref{fig:allv2}, our quark recombination model describes data rather well. In particular, the calculated elliptic flow $v_2$ exhibits a rapid rise at low $p_T$ and gradually approaches a saturation plateau in the intermediate $p_T$ region. This feature shows good consistency with the experimental measurements.

From the results listed in Tab.\ \ref{tab:fit}, it can be seen that the strange quark freeze-out proper time $\tau$ for 10–40\% centrality Au+Au collisions is 6.7-6.9 fm/$c$ when the collision energy increases from $\sqrt{s_{NN}}$ = 11.5 GeV to 62.4 GeV. This observation indicates that the volume of the fireball at hadronization has very weak dependence on the collision energy in this energy interval. We find the proper time $\tau$ reduces to 6.3 fm/$c$ at 7.7 GeV, which implies that the volume of the fireball begins to decrease. The ratio of event plane radii $R_y/R_x$ for Au+Au collisions at $\sqrt{s_{NN}}$ = 7.7–62.4 GeV is roughly constrained within the range of $1.24\pm0.04$, which also exhibits little dependence on the collision energy. The underlying physical reason is that the ratio $R_y/R_x$ is primarily governed by the initial spatial eccentricity of the overlapping collision zone. For events selected at the same collision centrality 10-40\%, this ratio should take very similar values across different collision energies. 

As for different centrality bins, both $\tau$ and $R_y/R_x$ show clear systematic trends: the value of $\tau$ decreases from central to peripheral collisions, reflecting the reduction of hadronization fireball volume and consistent with particle yield trends, while $R_y/R_x$ increases monotonically, reflecting the growing initial spatial eccentricity of the participant zone in more peripheral collisions.

\begin{figure}[tbh]
	\centering
	\begin{minipage}{0.36 \textwidth}
		\includegraphics[width=1\textwidth]{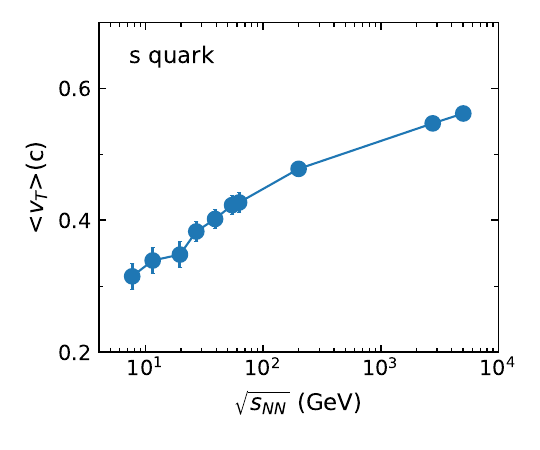}
	\end{minipage}
	\caption{\label{fig:flow}
		The averaged transverse flow velocity of strange quarks at hadronization: 10–40\%, 0–30\%, and 10–20\% centrality at $\sqrt{s_{NN}}$=7.7–62.4 GeV, 200 GeV, and 2.76-5.02 TeV, respectively.}	
\end{figure}

To characterize the behavior of the flow field, we calculate the averaged transverse flow velocity of strange quark at hadronization, which follows the relation $\left<v_T\right>=\frac{\int v_T  d^2r}{\int d^2r}=\frac{2}{3}~\alpha_0$. The corresponding results are presented in Fig.\ \ref{fig:flow}. For a fixed centrality bin, we find that the average velocity $\left<v_T\right>$ increases with collision energy, since a higher collision energy generates a larger pressure gradient that drives the fireball to expand more rapidly. We note that these systematic trends of $\tau$, $R_y/R_x$ and $\left<v_T\right>$ provide an important qualitative consistency check on the fit results and further validate the reliability of our calculations.  

\begin{figure}[tbh]
	\centering
	\begin{minipage}{0.48 \textwidth}
		\includegraphics[width=1\textwidth]{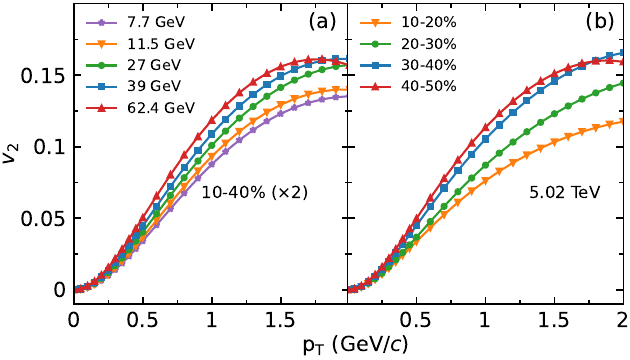}
	\end{minipage}
	\caption{\label{fig:v2sq}
		Elliptic flow $v_2$ of strange quarks at hadronization for selected collision energies. For ease of comparison, the values in panel (a) are scaled by a factor of 2.}	
\end{figure}

Using fit parameters given in Tab.\ \ref{tab:paraset} and Tab.\ \ref{tab:fit}, we can calculate the elliptic flow $v_2$ of strange quark at hadronization. Fig.\ \ref{fig:v2sq} shows the energy dependence (a) and centrality dependence (b) for the elliptic flow $v_2$ of strange quark at hadronization. In 10–40\% centrality Au+Au collisions, the $v_2$ curves for 19.6 GeV and 54.4 GeV are nearly identical to those of 39 GeV and 62.4 GeV, respectively, and are therefore not presented in Fig.\ \ref{fig:v2sq}(a). In general, we observe that the $v_2$ of strange quarks gradually increases with collision energy, growing by about 30\% as the collision energy rises from 7.7 GeV to 62.4 GeV, indicating that the collective flow behavior of strange quarks remains similar across this energy range. For Pb+Pb collisions at $\sqrt{s_{NN}}$ = 5.02 TeV, our calculations show that the $v_2$ of strange quark displays a clear increasing trend from central to peripheral collisions, which is fully consistent with the trend observed for the $v_2$ of $\phi$ and $\Omega$ hadrons.

\begin{figure*}[tbh]
	\centering
	\includegraphics[width=0.97\textwidth]{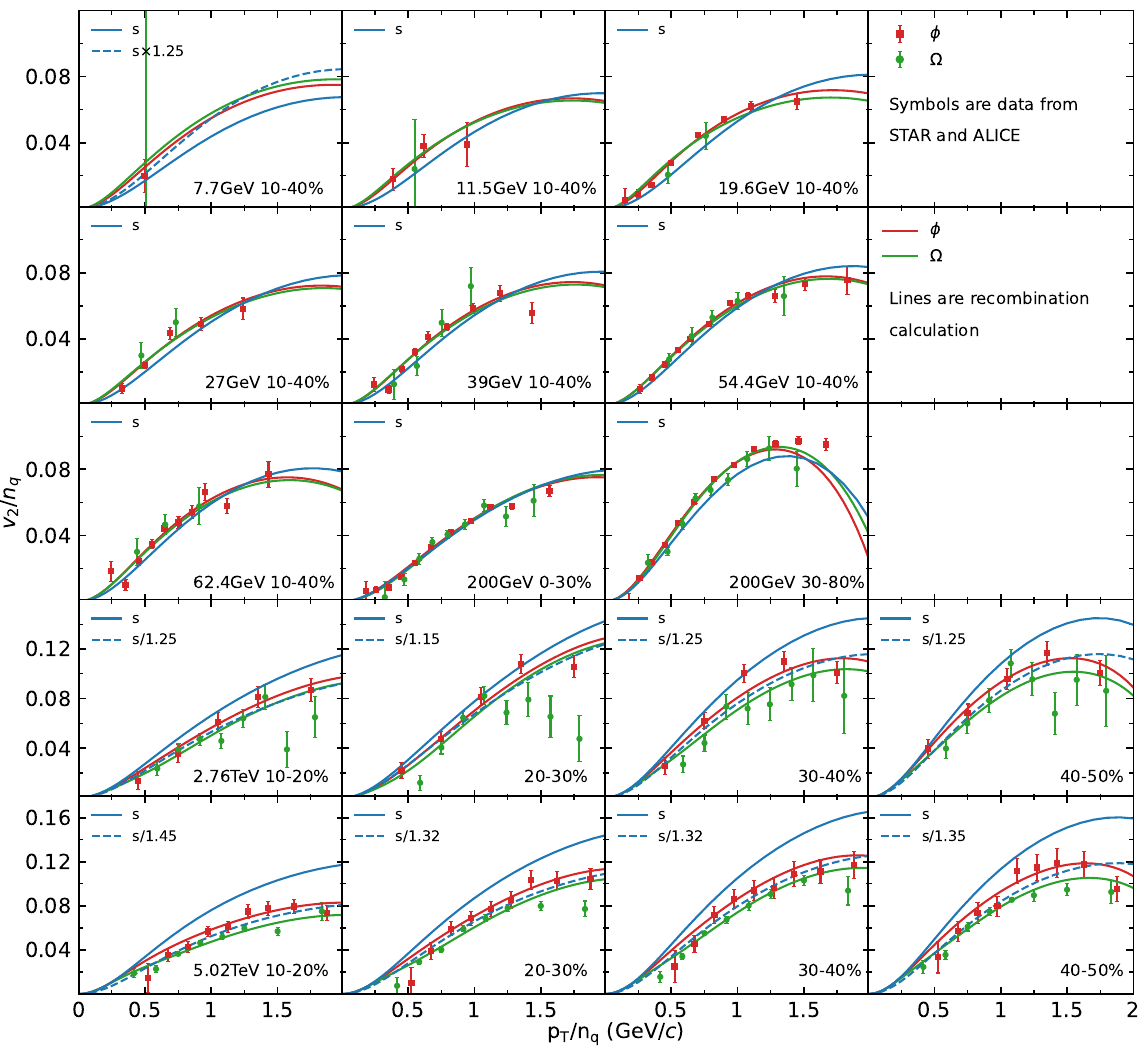}
	\caption{\label{fig:scaling}
 	   $v_2/n_q$ as a function of $p_T/n_q$ for $\phi$ mesons and $\Omega$ baryons at mid-rapidity in Au+Au collisions at $\sqrt{s_{NN}}$ = 7.7-200 GeV and Pb+Pb collisions at $\sqrt{s_{NN}}$ = 2.76-5.02 TeV. The $v_2$ for strange quark (labeled 's') are also shown. Lines are recombination calculation, symbols are data, details see text.}	
\end{figure*}

To explicitly exhibit the number-of-constituent-quark (NCQ) scaling property of the elliptic flow $v_2$ for $\phi$ and $\Omega$, we reprocess the data shown in Fig.\ref{fig:allv2} and construct a new plot in Fig.\ \ref{fig:scaling}, where the elliptic flow is scaled as $v_2/n_q$ and the transverse momentum is scaled as $p_T/n_q$, i.e. both are divided by number of constituent quarks $n_q$, with $n_q$ = 2 for the $\phi$ meson and $n_q$ = 3 for the $\Omega$ baryon. The results in Fig.\ \ref{fig:scaling} show that the calculated NCQ-scaled $v_2$ of $\phi$ and $\Omega$ are approximately equal across the collision energy range from 7.7 GeV to 5.02 TeV, which is fully consistent with the existing experimental measurements. Especially, a nearly perfect NCQ scaling is observed for $\phi$ and $\Omega$ in Au+Au collisions at $\sqrt{s_{NN}}$ = 7.7–200 GeV. We notice that for Pb+Pb collisions at $\sqrt{s_{NN}}$ = 2.76–5.02 TeV, the scaled $v_2$ of the $\phi$ meson is slightly higher than that of the $\Omega$ baryon, with a mean relative enhancement of $\sim 10\%$ at 2.76 TeV and $\sim 15\%$ at 5.02 TeV.

We further compare the elliptic flow $v_2$ of strange quark together with the NCQ-scaled elliptic flow $v_2$ of $\phi$ meson and $\Omega$ baryon, as shown in Fig.\ \ref{fig:scaling}. We find that the $v_2$ of strange quarks is close to the NCQ-scaled $v_2$ of $\phi$ and $\Omega$ in Au+Au collisions at $\sqrt{s_{NN}}$ = 11.5-200 GeV, which is in good agreement with the prediction from the naive quark coalescence model \cite{Molnar:2003ff,Kolb:2004gi}. The $v_2$ of strange quark is noticeably smaller than the NCQ-scaled $v_2$ of $\phi$ and $\Omega$ at 7.7 GeV, which is most likely because the extracted parameters overestimate the elliptic flow of $\phi$ and $\Omega$. Note that the experimentally measured $v_2$ of $\phi$ and $\Omega$ at 7.7 GeV has very few data points, and the extracted parameters carry large uncertainties.

On the other side, we observe that the $v_2$ of strange quark is significant larger than the scaled $v_2$ of $\phi$ and $\Omega$ in Pb+Pb collisions at $\sqrt{s_{NN}}$ = 2.76-5.02 TeV. To quantify their difference, we rescale the original strange quark elliptic flow $v_2$ by a modified factor, such that the corrected $v_2$ approaches the NCQ-scaled $v_2$ of $\phi$ and $\Omega$. From Fig.\ \ref{fig:scaling}, we find that the modified factor takes a value of approximately 1.25 for Pb+Pb collisions at $\sqrt{s_{NN}}$ = 2.76 TeV, and around 1.35 for collisions at $\sqrt{s_{NN}}$ = 5.02 TeV, with slight discrepancies across different centrality bins. One interesting observation is that the correction factor takes a larger value at $\sqrt{s_{NN}}$ = 5.02 TeV compared to 2.76 TeV, which implies that the quark elliptic flow $v_2$ exhibits a stronger deviation from the NCQ scaling behavior at the higher LHC energy. We note that our results are in good agreement with previous investigations on the elliptic flow $v_2$ of strange quarks, which were performed within different quark coalescence formalisms \cite{Chen:2006vc,Pu:2018eei}. 

It should be pointed out that the elliptic flow $v_2$ of strange quarks is equal to the scaled $v_2$ of $\phi$ and $\Omega$ only in the naive momentum-space quark recombination model, in which space-momentum correlations are neglected and quarks are with equal momentum \cite{Molnar:2003ff,Kolb:2004gi}. These assumptions hold only in the regime of low quark phase-space density, where the quark coalescence processes are relatively rare \cite{Molnar:2003ff}. At high phase-space density of constituent quarks, the momentum-space correlations may become significant and lead to the suppression of hadron elliptic flows, thus breaking the preconditions of the naive quark recombination model. Ref.\ \cite{Singha:2016aim} has investigated the NCQ scaling for the elliptic flow $v_2$ of identified hadrons at top-RHIC and LHC energies under the framework of A Multi-Phase Transport model (AMPT). They found that the observed breaking of NCQ scaling between mesons and baryons originates from the high phase-space density of constituent quarks in Pb+Pb collisions at $\sqrt{s_{NN}}$ = 2.76 TeV \cite{Singha:2016aim}, and the same mechanism may also account for the deviation of the strange-quark $v_2$ from the scaled $v_2$ of $\phi$ and $\Omega$.

\begin{figure}[tb]
	\centering
	\begin{minipage}{0.38 \textwidth}
		\includegraphics[width=1\textwidth]{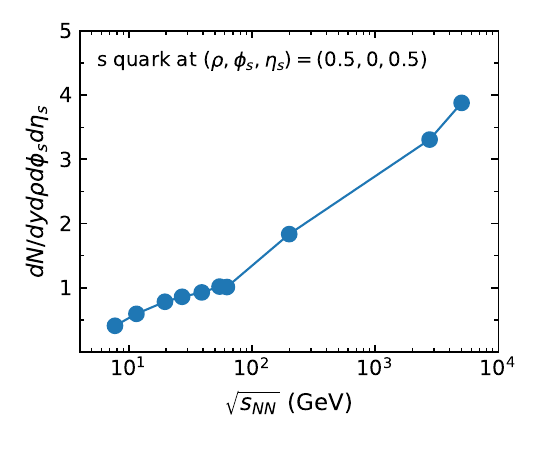}
	\end{minipage}
	\caption{\label{fig:density}
		Strange quark number density at mid-rapidity, evaluated at $\rho=0.5$, $\phi_s=0$ and $\eta_s=0.5$. For $\sqrt{s_{NN}}=200$ GeV we calculate 0-30\% centrality, and for $\sqrt{s_{NN}}$= 2.76-5.02 TeV we calculate 10-20\% centrality.} 
\end{figure}

To directly verify this statement, we calculate the strange quark number density using the quark distribution given in Eq.\ (\ref{eq:fdist}), by integrating over momentum space, thereby obtaining the density as a function of position. For illustration, we evaluate the density at $(\rho,\phi_s,\eta_s)=(0.5,0,0.5)$ (the parameters are defined in Sec.\ \ref{subsec:model}) for different collision energies, as shown in Fig.\ \ref{fig:density}. We find that the strange quark number density increases slowly to 1 when the collision energy increases from $\sqrt{s_{NN}}$ = 7.7 GeV to 62.4 GeV, and then increases rapidly, reaching 2 at 200 GeV and 3.5 at 2.76–5.02 TeV. ​In other words, the strange quark number density at 2.76-5.02 TeV is approximately 3.5 times that at 7.7–62.4 GeV, and 1.8 times that at 200 GeV at the selected point. This provides a plausible explanation for our finding that the strange quark elliptic flow $v_2$ deviates from the NCQ scaling behavior in Pb+Pb collisions at LHC energies.

\section{Conclusion}
\label{sec:sum}

In summary, we introduce viscous corrections into quark recombination by parameterizing quark distribution with a viscous blastwave. Using the quark recombination, we show that the elliptic flow $v_2$ of $\phi$ mesons and $\Omega$ baryons in heavy-ion collisions can be describled by quark recombination in a wide energy range from $\sqrt{s_{NN}}$ = 7.7 GeV to 5.02 TeV. We check the number-of-constituent scaling properties of $v_2$ of $\phi$ and $\Omega$ and find the NCQ-scaled $v_2$ of $\phi$ and $\Omega$ are approximately equal across the collision energy range from 7.7 GeV to 5.02 TeV, although some deviations are observed at 2.76 TeV and 5.02 TeV. 

We further present the elliptic flow $v_2$ of strange quarks at hadronization extracted from $\phi$ and $\Omega$ data. We find the $v_2$ of strange quarks is close to the NCQ-scaled $v_2$ of $\phi$ and $\Omega$ at $\sqrt{s_{NN}}$ = 11.5-200 GeV. On the other side, we find that the elliptic flow $v_2$ of strange quarks is significantly larger than the NCQ-scaled $v_2$ of $\phi$ and $\Omega$ at $\sqrt{s_{NN}}$ = 2.76–5.02 TeV, with the strange quark $v_2$ being 1.25 times and 1.35 times the latter at 2.76 TeV and 5.02 TeV, respectively. Such a deviation from the NCQ scaling may originate from the high phase-space density of constituent quarks in Pb+Pb collisions at 2.76-5.02 TeV.

Our studies imply that the phase-space density of quarks play an important role in hadron anisotropic flows, and special caution is required when extracting the partonic elliptic flow from the measured hadronic elliptic flow of hadrons. Only under the condition of low quark phase-space density (which typically corresponds to lower collision energies) can the partonic elliptic flow be interpreted as the NCQ-scaled $v_2$ of hadrons, as predicted by the naive parton recombination model. For an accurate extraction of the partonic flow, a more realistic recombination framework is necessary.

Our results verify that the quark recombination mechanism is valid for $\phi$ and $\Omega$ production in heavy-ion collisions from $\sqrt{s_{NN}}$ = 7.7 GeV to 5.02 TeV, indicating the existence of constituent quark degrees of freedom at these energies. Notably, the breaking of NCQ scaling for identified hadron elliptic flow could be potentially useful as a signal of phase transition in low-energy heavy-ion collisions \cite{Tian:2009wg,Liu:2025eml,STAR:2025owm}. It is highly interesting to apply the model and method of this work to the RHIC beam energy scan (BES) program. These ongoing studies will be presented elsewhere.

\begin{acknowledgments}

The authors would like to thank Rainer J.\ Fries, Che-Ming Ko, Yifeng Sun, Shuai Y.\ F.\ Liu, Jie Pu and Kai-Jia Sun for useful discussions. This work was supported in part by the National Natural Science Foundation of China under Grant Nos. 12205182 and 12235010, and the National SKA Program of China No. 2020SKA0120300.

\end{acknowledgments}

\bibliography{references} 

\end{document}